\documentclass[twocolumn,aps,amsmath,amssymb]{revtex4-1}
\usepackage{graphicx,bm,amsmath,amssymb,natbib,xcolor}

\begin{document}

\title{From adiabatic to dispersive readout of quantum circuits}
\author{Sunghun Park\textsuperscript{1}, C. Metzger\textsuperscript{2}, L. Tosi\textsuperscript{2,3}, M. F. Goffman\textsuperscript{2}, C. Urbina\textsuperscript{2}, H. Pothier\textsuperscript{2}, and A. Levy Yeyati\textsuperscript{1,4}}
\email[Corresponding author~: ]{a.l.yeyati@uam.es}
\affiliation{\textsuperscript{1}Departamento de F\'{\i}sica Te\'orica de la Materia Condensada, Condensed Matter Physics Center (IFIMAC), Universidad Aut\'onoma de Madrid, Spain\\
\textsuperscript{2}Quantronics group, Service de Physique de l'\'Etat Condens\'e (CNRS,
UMR\ 3680), IRAMIS, CEA-Saclay, Universit\'e Paris-Saclay, 91191 Gif-sur-Yvette, France\\
\textsuperscript{3}Centro At\'omico Bariloche and Instituto Balseiro, CNEA, CONICET, 8400 San Carlos de Bariloche, R\'io Negro, Argentina\\
\textsuperscript{4}Instituto Nicol\'as Cabrera, Universidad Aut\'onoma de Madrid, Spain
}

\date{\today}

\begin{abstract}
Spectral properties of a quantum circuit are efficiently read out by monitoring the resonance frequency shift it induces in a microwave resonator coupled to it.
When the two systems are strongly detuned, theory attributes the shift to an effective resonator capacitance or inductance that depends on the quantum circuit state. At small detuning, the shift arises from the exchange of virtual photons, as described by the Jaynes-Cummings model. Here we present a theory bridging these two limits and illustrate, with several examples, its necessity for a general description of quantum circuits readout.
\end{abstract}
\maketitle

Circuit Quantum Electrodynamics (cQED) is at the heart of most advanced superconducting quantum technologies.
Different types of superconducting qubits can be strongly coupled to microwave resonators thus achieving regimes and phenomenena which cannot be reached within the realm of quantum optics \cite{Blais2020}. 
More recently, strong coupling between microwave resonators and a variety of other quantum systems not necessarily involving superconductors has been achieved \cite{Clerk2020}, extending further the realm of cQED.
In all these applications the measurement of the qubit or the hybrid device state is achieved by monitoring the resonator properties. Theoretically, two regimes have been approached using disconnected descriptions \cite{Johansson2006}: 
the {\it dispersive} regime, where the qubit-resonator detuning is larger than the coupling strength 
yet small enough to allow  the exchange of virtual photons, and the {\it adiabatic} regime, where the detuning is sufficiently large for  virtual processes to be strongly suppressed.
The dispersive regime, which describes level repulsion between those of the quantum circuit and of the resonator, is typically dealt with using a Jaynes-Cummings Hamiltonian within different levels of approximation \cite{Blais2004,Wallraff2004,Johansson2006,Zueco2009,Smith2016,Kohler2018,Ansari2019,Zhu2020}. In contrast, the adiabatic regime accounts for the renormalization of the resonator capacitance/inductance by the effective capacitance of the circuit, including its ``quantum capacitance'' \cite{Sillanpaa2005,Duty2005},  or its effective inductance \cite{Sillanpaa2004,Paila2009}, which modifies the resonator frequency \cite{Persson2010,Bell2012}.

However, there is no actual border between these two regimes which could justify a separate treatment,  as illustrated by recent experiments on hybrid cQED setups \cite{Tosi2019} that reveal features of both regimes for the same device. This situation claims for a unified description of quantum circuits readout, going beyond the standard Jaynes-Cummings model, which could be applied to different types of devices over a large range of parameters. 

In the present Letter we derive a general expression for the resonator frequency shift when coupled to a generic quantum circuit. This expression naturally interpolates between the adiabatic and the dispersive regimes, thus allowing to clarify their origin from the same coupling Hamiltonian. In addition our formalism is not restricted to the usual two-level approximations but any multilevel situation can be described on the same footing. We illustrate the importance of the different terms in our expression by analyzing well-known models like a short single channel superconducting weak link hosting Andreev states, the RF-SQUID 
and the Cooper pair box.

\begin{figure}[t!]
\includegraphics[width=1\columnwidth]{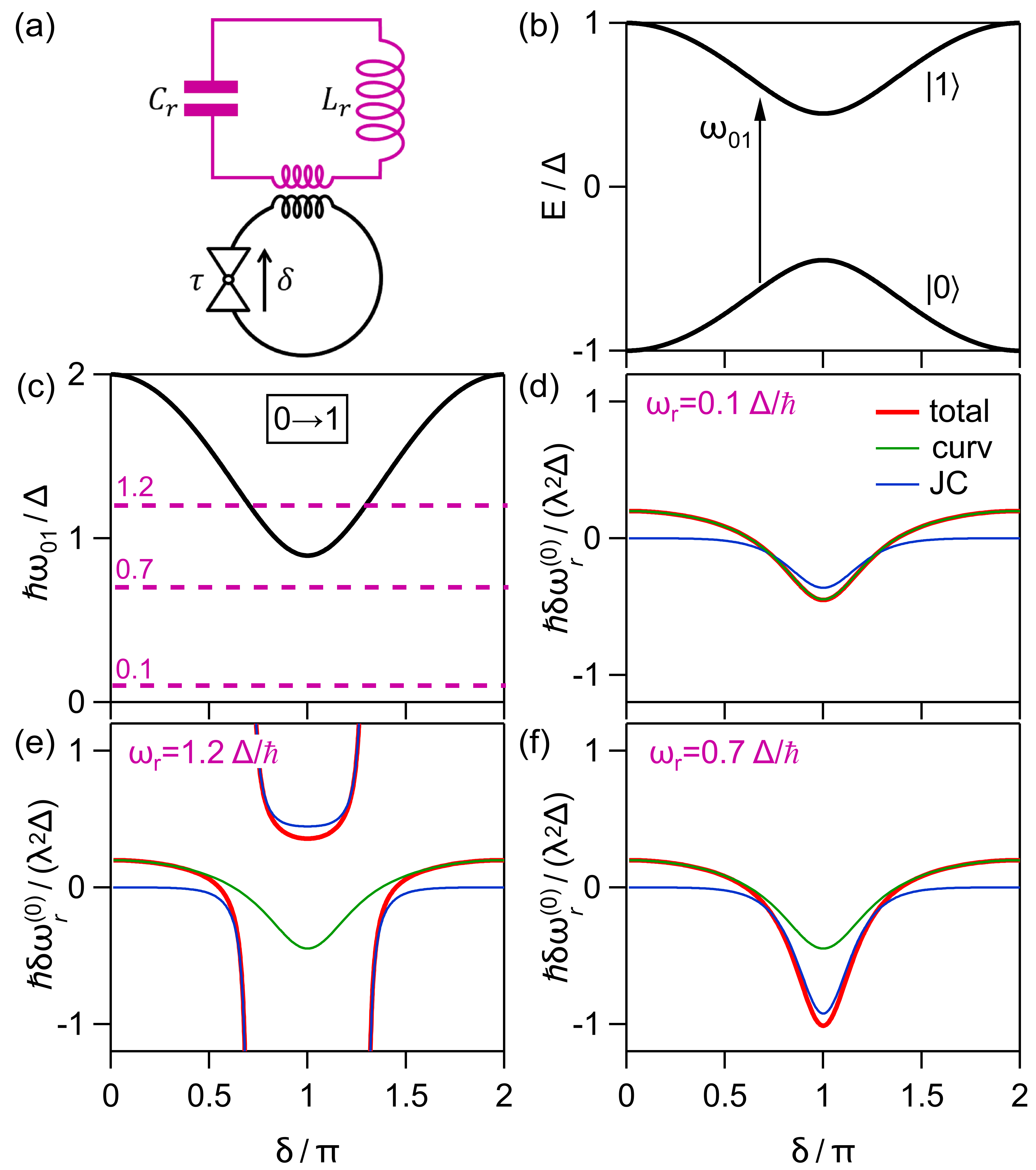}
\caption{Short single-channel weak link. (a) Circuit layout: the loop containing the phase-biased weak link of transmission $\tau$ is coupled to a microwave resonator (top). (b) Phase-dependence of the energy levels for $\tau=0.8$. (c) Transition energy $\hbar \omega_{01}=2E_A$; (d,e,f) resonator frequency shift in ground state $\delta \omega_r^{(0)}$ as a function of phase $\delta$ across weak link, for three values of the resonator frequency indicated with magenta dashed lines in (c). Red line: total shift; green line: curvature contribution; blue line: JC contribution (Eq.~(\ref{JC})).}
\label{Fig:ACshift}
\end{figure}

\begin{figure}[t!]
\includegraphics[width=1\columnwidth]{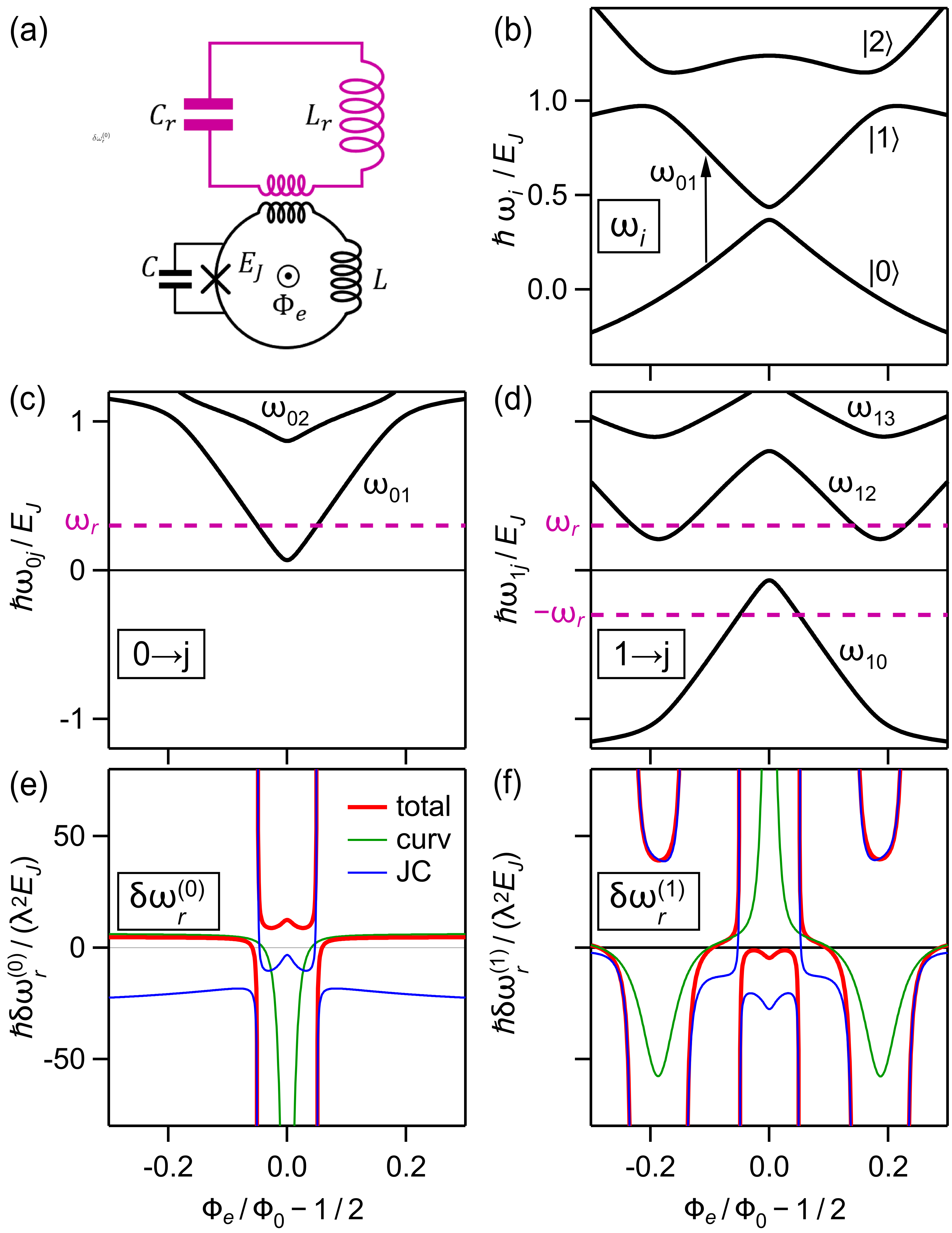}
\caption{RF-SQUID. (a) Schematics of the circuit, consisting of a Josephson junction with Josephson energy $E_J$ and capacitance $C$, inserted in a loop with inductance $L$ threaded by a magnetic flux $\Phi_e,$ and coupled to a microwave resonator (top). (b) Spectrum calculated using $E_C=E_L=E_J/5.$ (c,d) Transition energy $\omega_{0i}$ (resp. $\omega_{1i}$) from $|0\rangle$ (resp. $|1\rangle$). (e,f) Resonator frequency shift $\delta \omega_r^{(0)}$ (resp. $\delta \omega_r^{(1)}$) as a function of  $\Phi_e/\Phi_0-1/2, $ for resonator placed at $\omega_r=0.3E_J/\hbar$ (magenta dashed lines in (c,d)). Red line: total shift; blue line: JC contribution; green line: curvature contribution.}
\label{Fig:rfSQUID}
\end{figure}

\begin{figure}[t!]
\includegraphics[width=1\columnwidth]{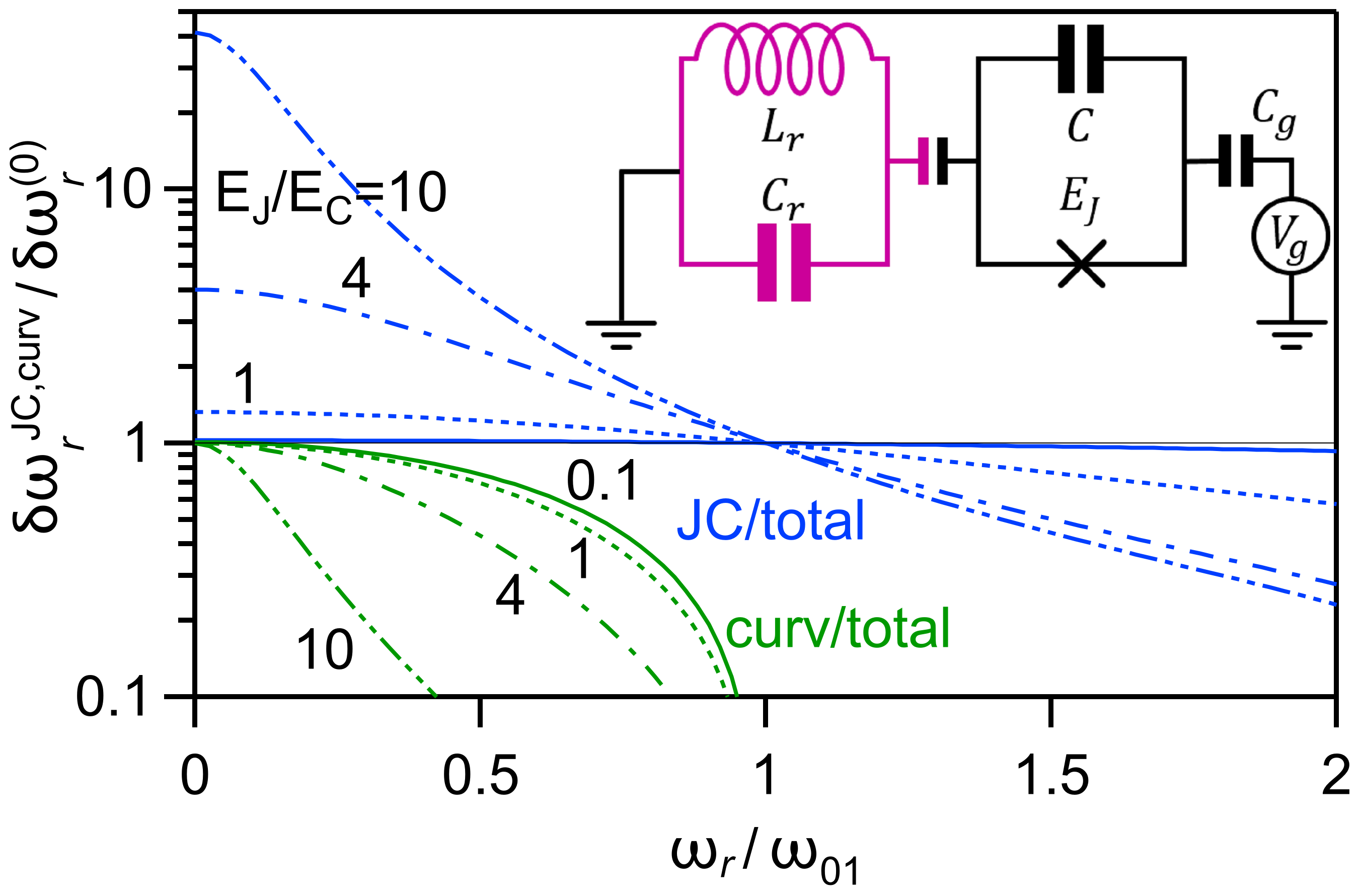}
\caption{Cooper pair box. Ratio of JC ($\delta \omega^{\text{JC}}_r$, blue) and curvature ($\delta \omega^{\text{curv}}_r=\lambda^2 \omega_0''$, green) contributions to total resonator frequency shift $\delta \omega_r^{(0)}$,  at $N_g=C_gV_g/2e=1/2$, for $E_J/E_C=0.1$, 1, 4 and 10, as a function of resonator frequency $\omega_r$ normalized to transition frequency $\omega_{01}.$ Inset: schematics of the Cooper pair box (black) coupled to microwave resonator (magenta).
}
\label{Fig:CPBshift}
\end{figure}

\textit {Resonator-quantum circuit coupling. ---}
The system we consider 
comprises a resonant circuit and a quantum circuit coupled through phase or charge fluctuations as depicted in Figs.~\ref{Fig:ACshift}(a), \ref{Fig:rfSQUID}(a) and in the inset of Fig.~\ref{Fig:CPBshift}. The resonant circuit is represented as a lumped-element LC resonator with bare resonance frequency $f_r=\omega_r/2\pi$, with $\omega_r=1/\sqrt{L_rC_r}$. Introducing the photon annihilation (creation) operators $a$ ($a^{\dagger}$), it can be described by the Hamiltonian $H_{r}=\hbar \omega_{r} a^{\dagger} a$. On the other hand, the quantum circuit Hamiltonian, $\hat{H}_{qc}(x)$, depends on a dimensionless control parameter $x$, corresponding to an excess charge on a capacitor or a flux through a loop. 
We denote by $|\Phi_{i}(x)\rangle$ the eigenstates of the uncoupled quantum circuit, $\hat{H}_{qc}(x) |\Phi_{i}(x)\rangle = E_{i}(x) |\Phi_{i}(x)\rangle$. Flux (charge) fluctuations in the resonator lead to $x \rightarrow x_0 + \hat{x}_r$, where  
$\hat{x}_r = \lambda (s\, a + s^*\, a^{\dagger})$ with a coupling constant $\lambda$, depending on a coupling scheme \cite{SM}, and $s=1$  $(-i)$.
We assume $\lambda \ll 1$ in accordance with experiments.  The resonator-quantum circuit coupling Hamiltonian $\hat{H}_{c}$ is obtained by expanding $\hat{H}_{qc}(x_0 + \hat{x}_r)$ up to second order in 
$\hat{x}_r$
\begin{equation}
\hat{H}_{c}(x_0) = \hat{x}_r \hat{H}'_{qc}(x_0) + \frac{\hat{x}^2_r}{2} \hat{H}''_{qc} (x_0), \label{Coupling}
\end{equation}
where the prime stands for the derivative with respect to $x$. The Hamiltonian describing resonator, quantum circuit and their coupling is therefore
\begin{multline}
\hat{H}=\hbar \omega_r a^{\dagger}a^{} + \hat{H}_{qc}(x_0)+\lambda  \hat{H}'_{qc}(x_0)\left(s\, a + s^*\, a^{\dagger}\right) \\ +\lambda^2 \hat{H}''_{qc}(x_0) (a^{\dagger}a^{}+1/2) , \label{Hamiltoniantotal}
\end{multline}
where terms $\lambda^2 a^{(\dagger)2}$ leading to corrections of order $\lambda^4$ have been neglected. When the quantum circuit is described as a two-level system and the terms involving ${H}''_{qc}$ in Eq. (\ref{Hamiltoniantotal}) are neglected, this model corresponds to the well-known Jaynes-Cummings Hamiltonian.  

We shall now evaluate the resonator shift for a given state $|\Phi_i\rangle$ in the quantum circuit using perturbation theory up to second order in $\lambda$.
The Hellmann-Feynman theorem establishes that $E'_{i}=\langle \Phi_{i}| \hat{H}'_{qc}| \Phi_{i}\rangle$. Taking the derivative on both sides gives 
\begin{equation}
E''_{i}=\langle \Phi'_{i}|\hat{H}'_{qc}|\Phi_{i}\rangle
+\langle \Phi_{i}|\hat{H}''_{qc}|\Phi_{i}\rangle 
+\langle \Phi_{i}|\hat{H}'_{qc}|\Phi'_{i}\rangle. \label{Epp}
\end{equation}
Here, $|\Phi'_{i}\rangle=\partial |\Phi_i\rangle/\partial x$ can be expressed as $|\Phi'_{i}\rangle = - G_i \left(G_i^{-1}\right)' |\Phi_{i}\rangle$ where $G_i = (E_i - \hat{H}_{qc})^{-1}$. Substituting this into Eq.~\eqref{Epp} and using identity $\sum_{i}|\Phi_{i}\rangle \langle\Phi_{i}|=1$, we obtain the relation between the diagonal matrix element of $\hat{H}''_{qc}$ and the curvature $E''_{i}$ of the energy level $i$, 
\begin{equation}
\langle \Phi_{i}| \hat{H}''_{qc}|\Phi_{i}\rangle=  E''_{i} +
2 \sum_{j\neq i} \frac{|\langle \Phi_{i}| \hat{H}'_{qc}|\Phi_{j}\rangle|^2}{E_{j}-E_{i}}. \label{Indt}
\end{equation}

Combining this result with the second order correction of the system energy levels arising from the $\hat{H}'_{qc}$ term in Eq. (\ref{Hamiltoniantotal}) \cite{SM}, we obtain the shift of the energy of the coupled system when the circuit is in state $|\Phi_i\rangle$ and the resonator contains $n$ photons
\begin{equation}
    \delta \omega_{i,n}=\left(n+\frac{1}{2}\right) \delta \omega_r^{(i)}+ \sum_{j \neq i} \frac{g_{i,j}^2}{2} \left(\frac{1}{\omega_{i j} -  \omega_r} - \frac{1}{\omega_{i j} + \omega_r} \right),\label{fShift-Level}
\end{equation}
where the shift $\delta \omega_r^{(i)}$ of the resonator frequency reads
\begin{equation}
\delta \omega_r^{(i)}=\lambda^2 \omega''_{i} + \sum_{j \neq i} g_{i,j}^2 \left(\frac{2}{\omega_{ij}}-\frac{1}{\omega_{ij} -  \omega_r} - \frac{1}{\omega_{i j} + \omega_r} \right),\label{fShift-Single}
\end{equation} 
with $\hbar g_{i,j}=\lambda |\langle \Phi_{i}| \hat{H}^{\prime}_{qc}|\Phi_{j}\rangle|$ the coupling strength between states $i$ and $j,$ $\omega_i=E_i/\hbar$ and $\omega_{i j}=\omega_j-\omega_i.$ Equations ~(\ref{fShift-Level},\ref{fShift-Single}) are the main results of this work, in particular Eq.~(\ref{fShift-Single}) contains both the adiabatic and the dispersive contributions to the resonator shift, as explained below. In the classical limit, it can be related to the real part of the AC current susceptibility as calculated in Ref.~\cite{Trif2018} for a fermionic system in thermal equilibrium.

The $\omega_r$-independent terms on the right-hand side of Eq.~(\ref{fShift-Single}) are the contributions involving $\hat{H}''_{qc}$ that arise from Eq.~\eqref{Indt}, while the $\omega_r$-dependent terms correspond to those obtained from a multi-level Jaynes-Cummings Hamiltonian.
It can be seen from Eq.~(\ref{fShift-Single}) that all transitions which couple a given state $i$ with other states $j$ via $\hat{H}'_{qc}$ are relevant to calculate the shift $\delta \omega_r^{(i)}$ of the resonance frequency. The equation includes the contribution from both, virtual transitions that do not depend on the resonator and other mediated by the absorption and emission of photons. Equation~(\ref{fShift-Single}) only holds far from resonances, \textit{i.e.} when all transitions between states of the circuit have energies that differ from $\omega_r$ by much more than the coupling energy. 

In the limit where $\omega_r \ll \omega_{i j}$ for all transitions, Eq.~(\ref{fShift-Single}) simplifies to $
\delta \omega_r^{(i)} \approx \delta \omega_r^\text{curv}= \lambda^2 \omega''_{i},
$ corresponding to a frequency shift proportional to the curvature of the energy level with $x$. Noting that for a charge-parameter $q$, $\left(\partial^2 E_{i}/\partial q^2\right)^{-1}$ is the effective capacitance \cite{Averin2003} of the circuit in state ${i}$, and for a phase-parameter $\varphi$, $(\Phi_0/2\pi)^2\left(\partial^2 E_{i}/\partial \varphi^2\right)^{-1}$ its effective inductance (here $\Phi_0$ is the flux quantum), this limit finds a simple interpretation: the resonator capacitance/inductance is merely renormalized by that of the quantum circuit.

It is only in the case where terms from $H''_{qc}$ are negligible that one recovers the result that can be derived from the generalized Jaynes-Cummings Hamiltonian \cite{Koch2007}, in which the frequency shift is dominated by the contributions involving the exchange of excitations
\begin{equation}
\delta \omega_r^{(i)}\approx \delta \omega_r^\text{JC}=-\sum_{j \neq i} g_{i,j}^2 \left(\frac{1}{\omega_{i j} -  \omega_r} + \frac{1}{\omega_{i j} + \omega_r} \right).
\label{JC}
\end{equation}
In the following, we will use the shortcut ``JC'' for this contribution.
For a quantum circuit described by a two-level system $\{|0\rangle, |1\rangle \}$, this result was derived from the Jaynes-Cummings Hamiltonian in the dispersive limit beyond the rotating-wave approximation (RWA) in Refs.~\cite{Johansson2006, Zueco2009}. Assuming $g_{01} \ll |\omega_{01}-\omega_r| \ll \omega_{01}+\omega_r$, it simplifies to $\delta \omega_r^{(0/1)} \sim \mp g_{01}^2/\left({\omega_{01} -\omega_r} \right)$, which is the cavity-pull $\chi_{01}$ in the RWA \cite{Blais2004}. When restricting to the three lowest energy levels of a multi-level circuit, Eq.~(\ref{fShift-Single}) also allows recovering the shifts derived for the Transmon in the RWA in Ref.~\cite{Koch2007}:
$\delta \omega_r^{(0)}\approx-\chi_{01}$, $\delta \omega_r^{(1)}\approx\chi_{01}-\chi_{12},$ and $\chi_{ij}=g_{ij}^2/(\omega_{ij}-\omega_r).$

Altogether, Eq.~(\ref{fShift-Single}) shows that the curvature of the energy levels, \textit{i.e.} the effective admittance of the circuit, is actually a distinct contribution to the shift and can be described on the same footing as the cavity pull given by the Jaynes-Cummings Hamiltonian. This result clarifies a link between both that had been suggested in early works \cite{Blais2004,Johansson2006}. 

\textit{Short weak link. ---} As a first example we address the case of a resonator inductively coupled to a small loop closed through a short, single-channel superconducting weak link. In a simplified low-energy description and neglecting the presence of excess quasiparticles, this circuit is characterized by two levels, at energies $\omega_0=-E_A/\hbar$ and $\omega_1=E_A/\hbar$, with $E_A=\Delta\sqrt{1-\tau \sin^2{(\delta/2})},$ the Andreev energy, $\Delta$ the superconducting gap, $\tau$ the channel transmission and $\delta$ the phase across the weak link \cite{Beenakker1991a, Furusaki1991, Bagwell1992}. The sole matrix element required to calculate the frequency shifts adopts the following analytical form \cite{Zazunov2014,Janvier2015}
\begin{equation}
\langle 0 | H'|1\rangle=\frac{\Delta\sqrt{1-\tau}}{2} \left(\frac{\Delta}{E_A}-\frac{E_A}{\Delta}\right).
\label{AC-matelem}
\end{equation}

We show in Fig.~\ref{Fig:ACshift} the phase dependence of the resonator frequency shift when the Andreev levels are in the ground state $\delta \omega_r^{(0)}$ 
for $\tau=0.8,$ and three values of the resonator frequency $\omega_r$. In Fig.~\ref{Fig:ACshift}(d), $\omega_r=0.1 \Delta/\hbar$ is much smaller than $\omega_{01}$ at all phases, and $\delta \omega_r^{(0)}$ is precisely given by the term associated to the curvature $\lambda^2 \omega_0^".$ In Fig.~\ref{Fig:ACshift}(f), $\omega_r=0.7 \Delta/\hbar$ approaches $\omega_{01}$ at $\delta \approx \pi$, so that the shift is in this region very close to the JC contribution, whereas further from $\pi$ it is given by the curvature. In Fig.~\ref{Fig:ACshift}(e), $\omega_r=1.2 \Delta/\hbar$ crosses $\omega_{01}$, and the characteristic anticrossing behavior in $\delta \omega_r^{(0)}$ can be observed, well described by JC. Away from $\delta \approx \pi$, the curvature once again takes over. While the short junction limit provides a simple analytical example to illustrate the crossover from the adiabatic to the dispersive regimes, a richer behavior, including finite length, parity, and spin-orbit effects \cite{Park2017,Tosi2019,Hays2019}, will be analyzed elsewhere \cite{Metzger2020}.

\textit{RF-SQUID.---}To illustrate our result from Eq. (\ref{fShift-Single}) in a multilevel situation, we now address the RF-SQUID, used in particular as a  simple flux qubit \cite{Friedman2000,Devoret2004}. Its Hamiltonian reads 
\begin{eqnarray}
    H&=&4E_C\hat{N}^2 + \frac{E_L}{2} \hat{\varphi}^2  - E_J \cos\left(\hat{\varphi} +  2\pi\frac{ \Phi_e}{\Phi_0}\right),
    \label{fluxquRFHam}
\end{eqnarray}
where $\hat{N}$ is the number of Cooper pairs having crossed the Josephson junction, $\hat{\varphi}$ the phase across the loop inductance, $E_C=e^2/2C$ the charging energy, $E_J$ the Josephson energy and  $E_L=\Phi_0^2/4\pi^2 L$ the magnetic energy associated with the loop geometric inductance $L$. The external flux $\Phi_e$ threading the loop is the control parameter.
By numerical diagonalization of the Hamiltonian, we obtain the spectrum, shown in  Fig.~\ref{Fig:rfSQUID}(b) for  $E_L=E_C=E_J/5$, and the transition energies $\omega_{0j}$ from state $|0\rangle$ (c) and $\omega_{1j}$ from state $|1\rangle$ (d). The resonator frequency shifts $\delta \omega_r^{(0,1)}$ when the circuit is in $|0\rangle$ or $|1\rangle$ are shown in (e) and (f), for a resonator at $\omega_r=0.3 E_J/\hbar.$ The curvature and JC contributions are shown as green and blue lines, respectively. It is only close to the crossings $\omega_{01}\approx \omega_r$ (resp. $\omega_{12}\approx \omega_r$) that $\delta \omega_r^{(0,1)}$ coincide with the JC contribution. When $\omega_r\ll \omega_{01}$ (resp. $\omega_r\ll \omega_{12},\omega_{01}$), the contribution from the curvature almost coincides with the total shift. When none of these conditions is met, the complete formula is necessary to describe the frequency shift, as clearly seen in Figure~\ref{Fig:rfSQUID}(e,f). However, if $8\lambda^2E_C \ll \hbar\omega_{01},$, the JC expression is almost correct if one uses an effective resonator frequency $\hbar{\omega_r^{\rm eff}}=\hbar\omega_r+8\lambda^2E_C$.

\textit{Cooper Pair Box. ---}
\label{comparison-CPB}
We now consider the Cooper pair box, a circuit that has both been discussed from the adiabatic \cite{Sillanpaa2005,Duty2005, Paila2009} and from the dispersive \cite{Blais2004,Koch2007} point of view. Its Hamiltonian reads
\begin{eqnarray}
    H_{cpb}&=&4E_C (\hat N-N_g)^2- E_J \cos{\hat \varphi}, 
    \label{CPBHamil}
\end{eqnarray}
with now $\hat \varphi$ the phase across the Josephson junction, conjugated to $\hat N$, and $N_g$ the reduced gate voltage (see inset in Fig.~\ref{Fig:CPBshift}).

In this case, $\lambda^2H_{cpb}''=8\lambda^2E_C$ is a constant which, when added to the JC contribution, leads to the $N_g-$dependent shift $\lambda^2 \omega_i''$ in the limit $\omega_r \rightarrow 0$. 
This is illustrated in Fig.~\ref{Fig:CPBshift}, showing the ratio of the JC and curvature contribution to the total frequency shift as a function of $\omega_r$, when the circuit is in state $|0\rangle$, at $N_g=1/2$ and for various values of $E_J/E_C.$ 
At $\omega_r \ll \omega_{01},$ the JC result overestimates by far the shift, and $\delta \omega_r^{(0)}$ is given by the curvature (adiabatic regime): $\delta \omega_r^{\rm curv}/\delta \omega_r^{(0)}\approx 1.$ Around the anticrossing at $\omega_r = \omega_{01},$ the JC contribution becomes very large, so that the constant contribution $8\lambda^2E_C$ is relatively negligible (dispersive regime): $\delta \omega_r^{\rm JC}/\delta \omega_r^{(0)}\approx 1$. When $E_J/E_C\lesssim 1,$ the JC result is very close to the exact result for all resonator frequencies. In contrast, for $E_J/E_C \gg 1,$ the limiting expressions $\delta \omega_r^{\rm curv}$ and $\delta \omega_r^{\rm JC}$ are only valid at $\omega_r \approx 0$ and $\omega_r \approx \omega_{01},$ respectively. Between these two limits the complete expression of Eq. (\ref{fShift-Single}) is needed to account for the frequency shift.

\textit{Conclusion and outlook ---} 
We have introduced a formulation of cQED readout bridging between the adiabatic and the dispersive limits that have been used to describe the coupling of a quantum circuit to a resonator in different regimes.  
While we have 
illustrated our work
by considering simple models, it provides a means to describe quantitatively cQED experiments which explore large ranges of transition frequencies \cite{Zhu2020,Tosi2019}. In particular, this is of importance for the spectroscopy of mesoscopic systems, like quantum hybrid devices combining spin-active materials (strong spin-orbit semiconducting nanowires or two-dimensional electron gases, topological insulators, etc.) and superconductors, currently explored in quest of topological superconductivity.

\acknowledgments  
We thank our colleagues from the Quantronics group for useful discussions. 
This work has been supported by ANR contract JETS, by FET-Open contract AndQC, 
by the Spanish MINECO through Grant No.~FIS2014-55486-P, ~FIS2017-84860-R 
and through the ``Mar\'{\i}a de Maeztu'' Programme for Units of Excellence in R\&D (MDM-2014-0377). S. Park acknowledges support by ``Doctor Banco Santander-Mar\'{\i}a de Maeztu'' program. 
L.Tosi was supported by the Marie Sk\l{}odowska-Curie individual fellowship grant 705467, and C. Metzger by Region Ile-de-France in the framework of DIM SIRTEQ.

S.P., C.M. and L.T. contributed equally to this work.

\begin{widetext}

\section{Detailed derivation of Eqs. (5,6)}

To compute the second order correction of the energy levels of Hamiltonian (2) we introduce a basis set $\left\{ |\Phi_i n \rangle \equiv |\Phi_i\rangle \otimes |n \rangle \right\}$, where $|\Phi_i\rangle$ corresponds to the eigenstates of $\hat{H}_{qc}$ with eigenvalue $E_i$ and $|n\rangle$ to a state with $n$ photons in the resonator. Assuming that the $|\Phi_i\rangle$ states are non-degenerate the lowest order correction to the combined system energy levels can be written as $\delta E_{i,n} = \delta E^{(1)}_{i,n} + \delta E^{(2)}_{i,n}$ where

\begin{eqnarray}
\delta E^{(1)}_{i,n} &=& \lambda^2 \langle \Phi_i n | \hat{H}''_{qc} \left(a^{\dagger} a + \frac{1}{2}\right) |\Phi_i n \rangle = \lambda^2 \langle \Phi_i |\hat{H}''_{qc}| \Phi_i \rangle \left(n+\frac{1}{2}\right) \nonumber\\
\delta E^{(2)}_{i,n} &=& -\lambda^2 \sum_{j\ne i,n'} \frac{|\langle \Phi_j n'|\hat{H}'_{qc}\left(s\, a + s^* \,a^{\dagger}\right)|\Phi_i n\rangle|^2}{E_j + \hbar\omega_R(n'-n) - E_i}
\nonumber \\
&=& -\lambda^2 \sum_{j\ne i} |\langle \Phi_j|\hat{H}'_{qc}|\Phi_i\rangle|^2\left(\frac{n+1}{E_j + \hbar\omega_r - E_i} + \frac{n}{E_j - \hbar\omega_r - E_i}\right) \;.
\end{eqnarray}

The more compact expressions given by Eqs. (5) and (6) in the main text are obtained by substituting $\langle \Phi_i |\hat{H}''_{qc}|\Phi_i \rangle$ using Eq. (4), leading to 

\begin{equation}
\delta E_{i,n} = \lambda^2\left\{E''_i \left(n+\frac{1}{2}\right) - \sum_{j\ne i}
|\langle \Phi_j|\hat{H}'_{qc}|\Phi_i\rangle|^2\left(\frac{n+1}{E_j + \hbar\omega_r - E_i} + \frac{n}{E_j - \hbar\omega_r - E_i} - \frac{2n+1}{E_j - E_i} \right)\right\} \;,
\end{equation}

or, equivalently as
\begin{equation}
\delta E_{i,n} =  \hbar\delta\omega_{r,i} \left(n+\frac{1}{2}\right) -
\frac{\lambda^2}{2} \sum_{j\ne i} |\langle \Phi_j|\hat{H}'_{qc}|\Phi_i\rangle|^2
\left(\frac{1}{E_j+\hbar\omega_r-E_i}-\frac{1}{E_j-\hbar\omega_r-E_i}\right) \;,
\end{equation}
where
\begin{equation}
\hbar\delta\omega_{r,i} = \lambda^2\left\{ E''_i  - \sum_{j\ne i}
|\langle \Phi_j|\hat{H}'_{qc}|\Phi_i\rangle|^2\left(\frac{1}{E_j + \hbar\omega_r - E_i} + \frac{1}{E_j - \hbar\omega_r - E_i} - \frac{2}{E_j - E_i} \right)\right\} \;.
\label{delta-omega}
\end{equation}

In the presence of degeneracy, the derivatives, $\hat{H}'_{qc}$ and $\hat{H}''_{qc}$, may, or may not, break the degeneracy. When degeneracy is conserved, for example, spin of Andreev levels in a weak link, this perturbation result remains valid. The validity of Eq. (4) can be seen by expressing the current matrix element in an alternative way as, 
\begin{equation}
\langle \Phi_i|\hat{H}'_{qc}|\Phi_j\rangle = E'_j \, \delta_{ij}+ (E_{j}-E_{i}) \langle \Phi_i|\Phi'_j \rangle,
\end{equation}
leading to 
\begin{equation}
\langle \Phi_{i}| \hat{H}''_{qc}|\Phi_{i}\rangle=  E''_{i} +
2 \sum_{j\neq i} \frac{|\langle \Phi_{i}| \hat{H}'_{qc}|\Phi_{j}\rangle|^2}{E_{j}-E_{i}}
=E''_{i} +
2 \sum_{j\neq i} 
(E_{j}-E_{i}) |\langle \Phi_i|\Phi'_j \rangle|^2,
\end{equation}
exhibiting no singular behavior in the degenerate case. If there exists a $g$-fold degeneracy (for example, orbital degeneracy) at energy $E = E_i$ with degenerate states, $\left\{ |\Phi_{ia} \rangle \right\}$ with $a = 1,2, ... ,g$, and if the states $|\Phi_{ia} \rangle$ do not diagonalize $\hat{H}'_{qc}$ and $\hat{H}''_{qc}$, we need to solve the following secular equation to obtain $\delta E^{(1)}_{i,n}$ in Eq. (S1), 

\begin{equation}
\text{Det} \left| \frac{\lambda^2}{2} \hat{M}_{i} (2 n+1) - \delta E^{(1)}_{i,n} \right| = 0,
\end{equation}
where $ \hat{M}_{i}$ is the $g \times g$ matrix whose elements are given by $ (\hat{M}_{i})_{a,b} = \langle \Phi_{ia} |\hat{H}''_{qc}| \Phi_{ib} \rangle$. 

Finally, the quantities $\delta \omega_{i,n}$, $\omega_{i,j}$ introduced in the main text are simply related to the ones in this supplemental material by $\delta\omega_{i,n} = \delta E_{i,n}/\hbar$ and $\omega_{ij}=(E_j-E_i)/\hbar$.

\section{Resonator-quantum circuit coupling}

\begin{figure}[t]
\includegraphics[scale=0.3]{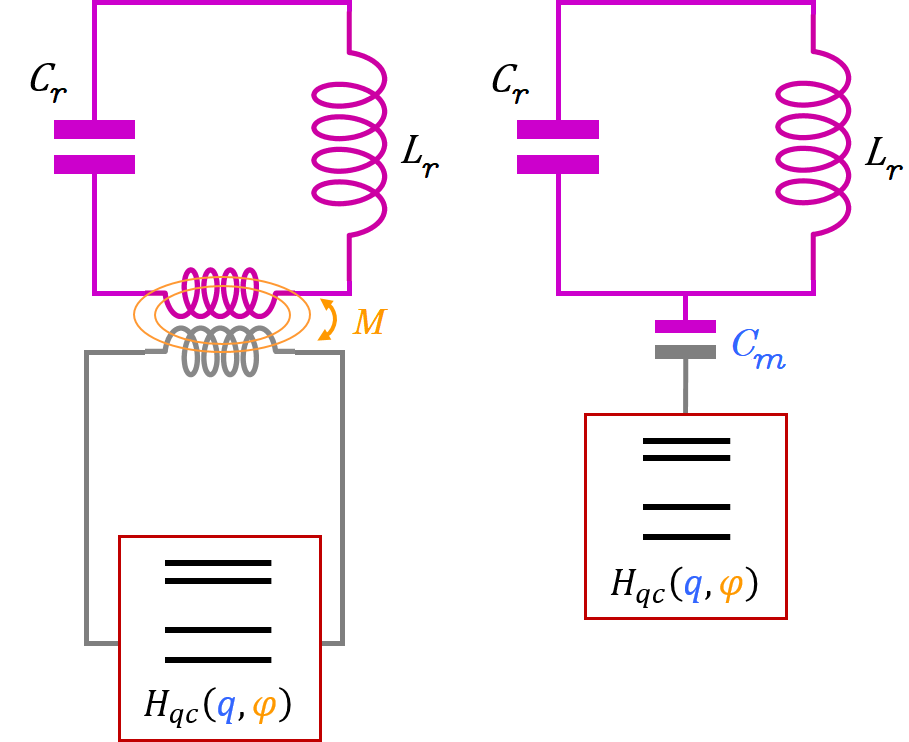}
\caption{Resonator-quantum circuit coupling schemes. Left: inductive coupling through a mutual inductance $M$. Right: capacitive coupling through capacitance $C_m$.}
\label{Fig:FigcQED+QS}
\end{figure}

We consider the general case of a quantum circuit coupled to a resonator either through a mutual inductance $M$ or through a coupling capacitance $C_m$ (Fig.~\ref{Fig:FigcQED+QS}). We express here explicitly the dimensionless coupling parameter $\lambda$. In the case of a capacitive coupling, $\lambda$ is the product of the geometric coupling ratio $C_m/C_r$ and of the reduced zero-point fluctuations of the charge $Q_r^{\text{zpf}}/2e$. One obtains
\begin{equation}
    \lambda=\frac{C_m}{C_r}\sqrt{\frac{R_Q}{4\pi Z_0}}
\end{equation}
with $Z_0=\sqrt{L_r/C_r}$ the resonator characteristic impedance, $R_Q=h/4e^2$ the resistance quantum.
In the case of an inductive coupling, $\lambda$ is the product of the geometric coupling ratio $M/L_r$ and of the reduced zero-point fluctuations of the phase $2\pi\Phi_r^{\text{zpf}}/\Phi_0$
\begin{equation}
    \lambda=\frac{M}{L_r}\sqrt{\frac{\pi Z_0}{R_Q}}.
\end{equation}

\section{Cooper pair box}

The Hamiltonian of the CPB is written in the charge basis as
\begin{eqnarray*}
    H&=&4E_C \sum_{N} (N-N_g)^2 |N\rangle \langle N| - \frac{E_J}{2} \sum_{N}\left(|N\rangle \langle N+1|+|N+1\rangle \langle N|\right)
\end{eqnarray*}
so that 
\begin{eqnarray}
    H'&=&-8E_C \sum_{N} (N-N_g) |N\rangle \langle N|
\end{eqnarray}
and
\begin{eqnarray}
    H''&=&8E_C.
    \label{Eq:H"Box}
\end{eqnarray}

\begin{figure}[h!]
\includegraphics[width=1\columnwidth]{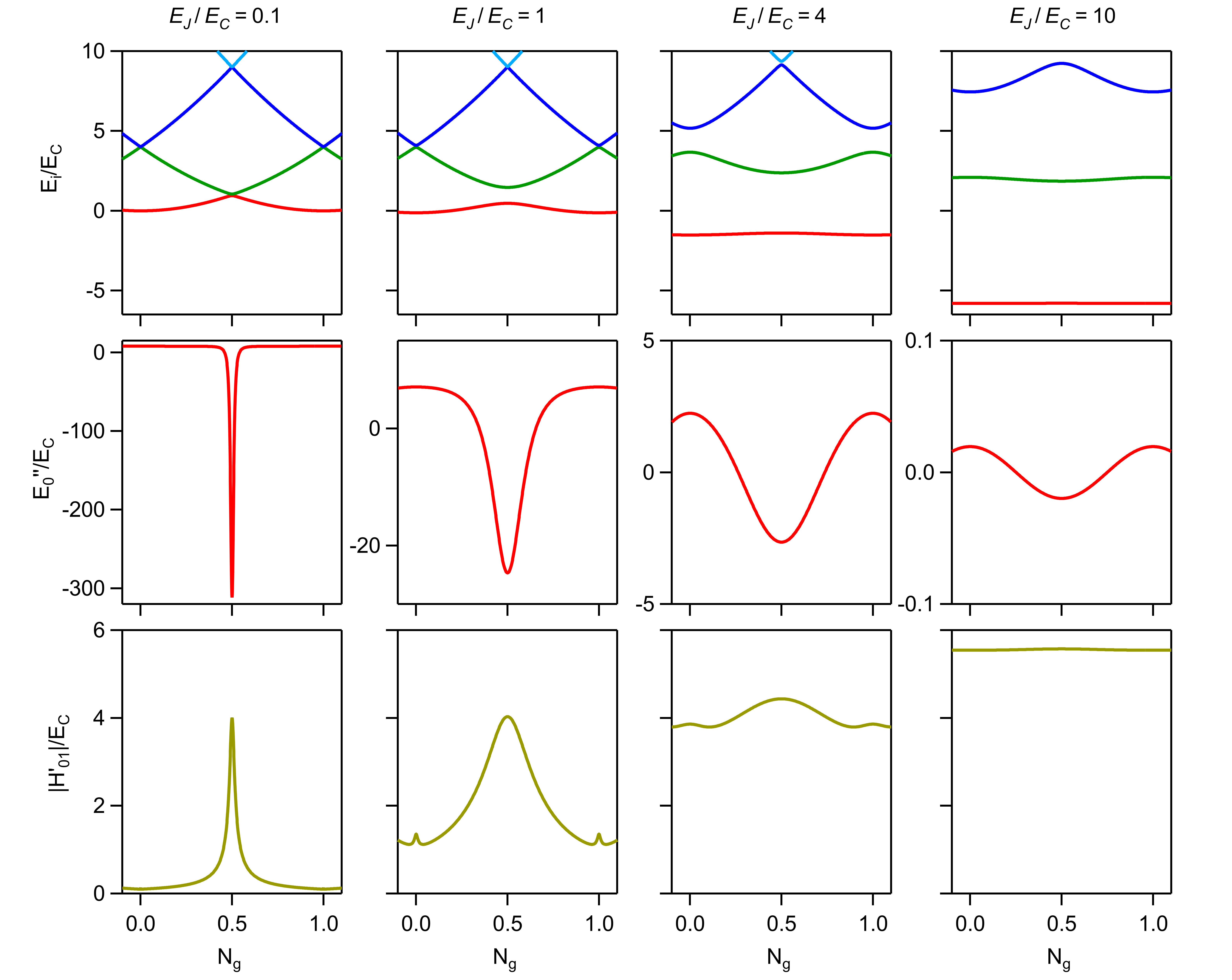}
\caption{Cooper pair box for $E_J/E_C=$0.1,1, 4  and 10 from left to right column. (top row) Spectrum. (mid row) Curvature for the ground state.  (bottom row) Matrix elements  $H'_{01}=\langle 0|H'|1\rangle$.}
\label{Fig:CPBsup}
\end{figure}


\subsection{Explicit derivation of Eq.~(4) in the two-level approximation}

In the case of very small $E_J/E_C$, one can work in the base of the two lowest charge states 
 $\{ |N=0\rangle, |N=1\rangle \}$ and resolve analytically the effective Hamiltonian
\begin{equation*}
    \tilde{H}=\left( \begin{array}{cc} 4E_C N_g^2 & - \frac{E_J}{2} \\
    - \frac{E_J}{2} & 4E_C(1-N_g)^2 \end{array}\right),   
\end{equation*}
with eigenvalues $E_{0,1}=2 E_C(2N_g^2+1-2N_g) \pm \sqrt{4E_C^2 (2N_g-1)^2 + (E_J/2)^2}$ and eigenvectors
\begin{eqnarray*}
    |0\rangle &=&u |N=0\rangle + v |N=1\rangle \\
    |1\rangle &=&-v |N=0\rangle + u |N=1\rangle,   
\end{eqnarray*}
where
\begin{equation*}
    u^2=\frac{1}{2} + \frac{4E_C(1-2N_g)}{4\sqrt{4E_C^2 (2N_g-1)^2 + (E_J/2)^2}},
\end{equation*}
and $v^2=1-u^2$. One obtains for the curvature 
\begin{equation}
    \frac{\partial^2E_{0,1}}{\partial N_g^2}=8E_C \pm \frac{4 E_C^2 E_J^2}{[4 E_C^2(1-2N_g)^2 + (E_J/2)^2]^{3/2}}.
    \label{Eq:curvBox}
\end{equation}

The matrix element between the eigenvectors is
\begin{eqnarray*}
    \langle0|H'|1\rangle &=& -8E_C uv\\
    \Rightarrow |\langle0|H'|1\rangle|^2&=&  \frac{4E_C^2E_J^2}{[4E_C^2 (2N_g-1)^2 + (E_J/2)^2]},
\end{eqnarray*}
hence
\begin{eqnarray}
    \frac{2|\langle0|H'|1\rangle|^2}{E_1-E_0} &=& \frac{4E_C^2E_J^2}{[4E_C^2 (2N_g-1)^2 + (E_J/2)^2]^{3/2}}.
    \label{Eq:H'box}
\end{eqnarray}
Using Eqs.~(\ref{Eq:H"Box},\ref{Eq:curvBox},\ref{Eq:H'box}) one recovers the result of Eq.~(4)
\begin{equation}
    \frac{\partial^2E_{0}}{\partial N_g^2}=\langle0 |H'' |0\rangle + \frac{2|\langle0|H'|1\rangle|^2}{E_1-E_0}
\end{equation}
and 
\begin{equation}
    \frac{\partial^2E_{1}}{\partial N_g^2}=\langle1 |H'' |1\rangle + \frac{2|\langle0|H'|1\rangle|^2}{E_0-E_1}.
\end{equation}

\subsection{Calculation in the general case}
In the general case for arbitrary $E_J/E_C$ one has to compute numerically the energy levels and the matrix elements of $H'$ by exact diagonalization of Hamiltonian given in Eq.~(10). We consider a truncated base with $20$ charge states. When calculating the resonator frequency shift, we take in the summation the number of terms such that Eq.~(4) is verified.

In Fig. \ref{Fig:CPBsup}(a) we show the energy of the first levels as a functions of $N_g$ for $E_J/E_C=$0.1,1,4 and 10 as corresponds to Fig. 3 in main text. In (b) we show the curvature of the ground state for each case and in (c) the matrix element $|\langle 1|H'|0\rangle|$.

\section{RF-SQUID}

\begin{figure}[t]
\includegraphics[width=\columnwidth]{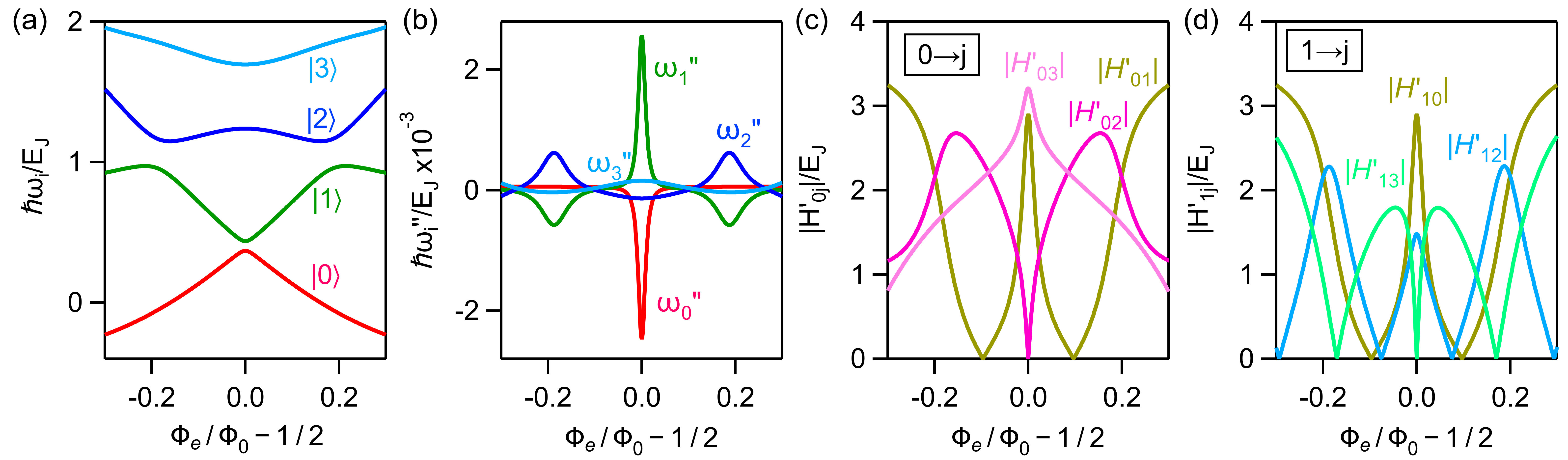}
\caption{RF-SQUID for $E_C=E_L=E_J/5$. (a) Spectrum. (b) Curvature for for the lowest energy states.  (c,d) Matrix elements  $H'_{0j}=\langle O|H'|j\rangle$  and  $H'_{1j}=\langle 1|H'|j\rangle$.}
\label{Fig:RFSQUIDsup}
\end{figure}

The Hamiltonian of the RF-SQUID can be written in the phase base as
\begin{eqnarray*}
    H&=&-4E_C\partial^2_\varphi + \frac{E_L}{2} \varphi^2  - E_J \cos\left(\varphi +  2\pi\frac{ \Phi_e}{\Phi_0}\right)=-4E_C\partial^2_\varphi +U(\varphi),
\end{eqnarray*}
which can be expressed as a tight-binding type Hamiltonian taking $\varphi=ma_0$, with $a_0$ the lattice parameter, $m$ an integer
\begin{eqnarray*}
    H_{tb}&=&\sum_{m} \left(\frac{8E_C}{a_0^2} + U(ka_0)\right) |m\rangle \langle m| -\frac{4E_C}{a_0^2} \sum_{m} \left( |m\rangle \langle m+1| + |m+1\rangle \langle m| \right).
\end{eqnarray*}

As in the case of the Cooper pair box this Hamiltonian is diagonalized numerically using $a_0=0.1$ and a matrix of 125x125 states. In Fig. \ref{Fig:RFSQUIDsup}(a) we show the energies of the first levels as a function of the external flux for $E_C=E_L=E_J/5$. The curvature of each state is shown in (b). In Fig. \ref{Fig:RFSQUIDsup}(c) and (d) we show resulting matrix elements $|H'_{0j}|=|\langle 0|H'|j\rangle|$ and $|H'_{1j}|=|\langle 1|H'|j\rangle|$, respectively.

\section{Short superconducting weak link}

We derive here explicitely Eq.~(4) for the short weak link, in the two-level approximation. In the simplified description we use in this Letter, the Hamiltonian and its derivatives can be expressed in the ``Andreev basis" in which $H_A$ is diagonal using the Pauli matrices $\hat{\sigma}_{x,y,z}$ (note however that the Andreev basis is itself phase-dependent)  \cite{Bretheau2013}:
\begin{align}
H_A = -E_A \hat{\sigma}_z,
\label{Eq:AndreevHam}
\end{align}
\begin{align}
H'_A =  \varphi_0 I_A\left(\hat{\sigma}_z+\sqrt{1-\tau}\tan(\delta/2) \hat{\sigma}_x\right),
\label{Eq:AndreevCurrent}
\end{align}
\begin{align}
H''_A = \varphi_0 I_A\left(
\frac{\tau + (2-\tau)\cos\delta}{2 \sin{\delta}} \hat{\sigma}_z-\sqrt{1-\tau}\hat{\sigma}_y\right).
\label{Eq:AndreevSecondDer}
\end{align}
where $\varphi_0=\Phi_0/2\pi$ and  $I_A=\Delta^2 \tau \sin{(\delta)}/ 4\varphi_0E_A$. Using these expressions, one checks easily that 
\begin{align}
\langle 0|H''_A|0\rangle = - E_A'' + 2 \frac{|\langle 0 | H'_A|1\rangle|^2}{2E_A}
\label{Eq:demo1}
\end{align}
and 
\begin{align}
\langle 1|H''_A|1\rangle = E_A'' + 2 \frac{|\langle 0 | H'_A|1\rangle|^2}{-2E_A}
\label{Eq:demo1}
\end{align}
as implied by Eq.~(4) in the main text.

\end{widetext}

\end{document}